\documentclass[aps,prd,groupedaddress]{revtex4}

\usepackage{amsmath,amssymb,bm}

\usepackage{graphicx}
\usepackage{amsfonts}
\usepackage{amssymb}
\usepackage{amsbsy}
\usepackage{amsmath}
\usepackage{latexsym}
\usepackage{bm}
\usepackage{color}
\usepackage{comment}

\newcommand*{\di}{\partial}

\renewcommand*{\c}{\text{c}}

\def\p {\partial}
\def\t {\tilde}
\def\be {\begin{eqnarray}}
\def\ee {\end{eqnarray}}
\def\nn {\nonumber}

\begin{document}

\title{Linearized 3D gravity with dust}

\author{Viqar Husain}
\email[]{vhusain@unb.ca}

\author{Shohreh Rahmati}
\email[]{srahmati@unb.ca}

\author{Jonathan Ziprick}
\email[]{jziprick@unb.ca}

\affiliation{University of New Brunswick\\
Department of Mathematics and Statistics\\
Fredericton, NB E3B 5A3, Canada}

\date{\today}

\begin{abstract}

Three-dimensional gravity coupled to pressureless dust  is a field theory with  one local  degree of freedom. In the canonical framework, the dust-time gauge encodes this physical degree of freedom as a metric function.  We find that the dynamics of this field, up to  spatial diffeomorphism flow, is  independent of spatial derivatives and is therefore ultralocal. We also derive the linearized equations about flat spacetime, and show that the physical degree of freedom may be viewed as either a traceless or a transverse mode. 

\end{abstract}

\maketitle

\section{Introduction}

Einstein gravity in three spacetime dimensions has been a subject of much study, primarily as a model for quantum gravity in a simpler setting \cite{Carlip:1994ap,Carlip:1998uc}. In vacuum however the theory has only a finite number of degree of freedom which arise  either due to  non-trivial topology of space \cite{Ashtekar:1989qd,Witten:1988,Moncrief:1990mk}, or through point particles which appear as conical defects \cite{Deser:1983tn}. It is therefore interesting to study three-dimensional gravitational theories  that do have local field degrees of freedom.  There are many such examples, prominent among them is the  topologically massive theory \cite{Deser:1981wh}.

We consider here another method for obtaining local degrees of freedom: three dimensional Einstein gravity coupled to matter. This has been studied before.  It is known for example that 3d gravity with a scalar field has wave solutions \cite{Husain:1994xa}. Here we study coupling to pressureless dust in the  Arnowitt-Misner-Deser (ADM) canonical framework.   We study the theory in the dust time gauge, where spatial slices are defined by level values of the scalar field. This  gauge has an interesting property: the (non-vanishing) physical Hamiltonian is the former Hamiltonian constraint \cite{Husain:2011tk}.   A counting of degrees of freedom reveals that the theory has one local (configuration) field degree of freedom, which in the dust-time gauge manifests itself as a metric field. In a previous work two of the present authors studied  the spherically symmetric sector of this theory \cite{Husain:2015cwa}. 

In this paper we investigate the nature of this field  without imposing additional symmetries, by analyzing the linearized theory. We begin by reviewing the canonical framework and the dust time gauge in the next section. We find that in the dust time gauge    space points decouple, and the dynamics at each point is independent and  identical, and also subject  to spatial diffeomorphism flow. In section 3 we analyze the linearized theory in Fourier space; we solve the diffeomorphism constraint and derive the linearized equations for the physical degree of freedom. We conclude in Section 4 with a brief summary and implications of the result for quantum gravity.

\section{Action and Hamiltonian theory}
 The theory  we study  is given by the action  
\be
S=\frac{1}{2\pi}\int{d^{3}x\sqrt{g}R}-\frac{1}{4\pi}\int{d^{3}x\sqrt{g}\ m(g^{\mu\nu}\di_{\mu}\phi\di_{\nu}\phi+1)}.
\ee
The first integral is the usual gravitational action with vanishing cosmological constant. The second integral is the action for the pressureless dust field, where $m$ is a function of spacetime.  Variation with respect to $m$ constrains the dust field to have a timelike gradient $|\nabla \phi|^2 = -1$.


The canonical ADM action is
\be
S=\frac{1}{2\pi}\int{d^{3}x\left(\tilde{\pi}^{ab}\dot{q}_{ab}+p_{\phi}\dot{\phi}-N\mathcal{H}-N^{a}\mathcal{C}_{a}\right)},
\ee
where  the pairs  $(q_{ab},\tilde{\pi}^{ab})$   and  $(\phi, p_{\phi})$ are respectively the gravitational and dust phase space variables.  The lapse and shift functions,  $N$ and $N^{a}$ are the coefficients of the Hamiltonian and diffeomorphism constraints
\be
\label{HG}
\mathcal{H} &=&\mathcal{H}^{G}+\mathcal{H}^{D},\\
\mathcal{C}_{a}&=&\mathcal{C}^{G}_{a}+\mathcal{C}^{D}_{a}=-2D_{b}\tilde{\pi}^{b}_{a}+p_{\phi}\di_{a}\phi,
\ee
where
\be
\mathcal{H}^{G}&=& \sqrt{q}\left(-R^{(2)}+\frac{1}{q}(\tilde{\pi}^{ab}\tilde{\pi}_{ab}-\tilde{\pi}^{2})\right),\\
\mathcal{H}^{D}&=&\frac{1}{2}\left(\frac{p_{\phi}^{2}}{m\sqrt{q}}+m\sqrt{q}(q^{ab}\di_{a}\phi\di_{b}\phi+1)\right).
\ee
The trace of the gravitational momentum is $\tilde{\pi}=q_{ab}\tilde{\pi}^{ab}$, $R^{(2)}$ is the scalar curvature of the spatial hypersurfaces, and $D_a$ is the covariant derivative associated with $q_{ab}$.

The momentum conjugate to the field $m$ is zero since it appears as a Lagrange multiplier in the covariant action. However, it is still present in   $\mathcal{H}^{D}$.  The  canonical action may be written in a convenient form that contains only the phase space variables by varying the action with respect to $m$ and substituting back the solution. This gives
\be
\label{m}
m=\pm\frac{p_{\phi}}{\sqrt{q(q^{ab}\di_{a}\phi\di_{b}\phi+1)}},
\ee
which leads to
\be
\mathcal{H}^{D}=  \pm\  p_\phi  \sqrt{q^{ab}\di_{a}\phi\di_{b}\phi+1}.
\ee
It is readily verified that the constraints remain  first class. We will see in the gauge fixing below how the sign is selected.

\subsection{Time gauge fixing}
We now partially reduce the theory  by fixing a time gauge and solving the Hamiltonian constraint to obtain a physical Hamiltonian. We use the dust time gauge  \cite{Husain:2011tk,Swiezewski:2013jza} which equates the physical time with level values of the scalar field:
\be
\label{gauge}
\lambda\equiv \phi-t\approx0.
\ee
This has a nonzero Poisson bracket with the Hamiltonian constraint, so this pair of constraints is second class. Requiring that the gauge condition be preserved in time gives an equation for the lapse function:
\be
1 = \dot{\phi}= \left. \left\{\phi, \int d^3x \left (N  \mathcal{H}  + N^a \mathcal{C}_{a}\right) \right\} \right|_{\phi=t}  =   N\  \frac{p_\phi}{m\sqrt{q}} \implies N= \frac{\sqrt{q}m}{p_\phi}.
\ee
Using the relation  $(\ref{m})$   with $\phi=t$   leads to  $N=\pm 1$ and implies that $\sqrt{q} m=\pm p_\phi$. The sign of the lapse function determines whether the evolution is forward ($N=+1$) or backward ($N=-1$) in time. We select the positive sign which fixes the above ambiguity in the Hamiltonian constaint, yielding $\mathcal{H}^{D}= +p_\phi$.

Imposing this gauge choice eliminates the Hamiltonian constraint from the theory. Imposing this constraint strongly gives 
\be
p_\phi = - \mathcal{H}^{G}.
\ee
Substituting this and the gauge condition (\ref{gauge}) into the canonical action gives
\be
S_{GF}=\frac{1}{2\pi}\int{d^{3}x\left(\tilde{\pi}^{ab}\dot{q}_{ab}-\mathcal{H}^{G} -N^{a}\mathcal{C}^{G}_{a}\right)},
\ee
This shows that the gravitational part of the Hamiltonian constraint becomes  the physical Hamiltonian, and the full diffeomorphism constraint  reduces to the gravitational one.  This is a field theory: 3 functions in $q_{ab}$ subject to the two diffeomorphism constraints gives one local configuration degree of freedom. This is the action we study in the remainder of the paper.

\subsection{Equations of motion}

The theory so far has been partially reduced using the dust time gauge. The equations of motion are obtained via Poisson brackets with the Hamiltonian
\be
\mathcal{H}=\frac{1}{2\pi}\int{d^{3}x\left(\mathcal{H}^{G}+N^{a}\mathcal{C}^{G}_{a}\right)} .
\ee
We have
\be
\dot{q}_{ab}=\{q_{ab},\mathcal{H}\}=\frac{2}{\sqrt{q}}\left(\tilde{\pi}_{ab}-\tilde{\pi}q_{ab}\right)+\mathcal{L}_{N}q_{ab} \label{3d1}
\ee
\be
\dot{\tilde{\pi}}^{ab}=\{\tilde{\pi}^{ab},\mathcal{H}\}= \frac{q^{ab}}{2\sqrt{q}}\left[\tilde{\pi}^{cd}\tilde{\pi}_{cd}-\tilde{\pi}^{2}\right]
- \frac{2}{\sqrt{q}} \left( \t{\pi}^a_c \t{\pi}^{cb} - \t{\pi} \t{\pi}^{ab} \right)
+\mathcal{L}_{N}\tilde{\pi}^{ab}, \label{3d2}
\ee
where $\mathcal{L}_N$ is a Lie derivative in the direction of the shift vector $N^a$.
Spatial derivatives enter these equations only through the Lie derivative terms, which is a gauge variation. There are no physically meaningful spatial derivatives because the Ricci scalar term $\sqrt{q} R^{(2)}$
does not contribute to the equations of motion. This can be seen by evaluating its variation:
\be
\delta \left( \sqrt{q} R^{(2)} \right) &=& \left( \delta \sqrt{q}\  \right) R^{(2)} + \sqrt{q} \left( R_{ab}^{(2)} \delta q^{ab}\right)  + \nabla_\c J^c  \nn\\
&=&  \sqrt{q}\left(  R_{ab}^{(2)} - \frac{R^{(2)}}{2} q_{ab}      \right) \delta q^{ab}  + \nabla_\c J^c \qquad \nn\\
&=&  \nabla_\c J^c,
\ee
where $\displaystyle J^c = q^{ab}\delta\Gamma^c_{ab} - q^{ca} \delta\Gamma^d_{ad}$, and the last equality follows  in $2-$dimensions from the formula for the Riemann tensor $\displaystyle R_{abcd} = (R/2) \left(q_{ac}q_{bd} -q_{ad} q_{bc}  \right)$. Since the variation is a total derivative, it contributes only to a boundary term and does not affect the dynamics.

We conclude from this that the dynamics is ultralocal: the evolution equations for the phase space variables contain no spatial derivatives in the non-gauge terms. This means that the degrees of freedom at each space point evolve independently  of any other point, and the diffeomorphisms serve only to move the points around.   We note however that if spatial coordinate gauges are fixed and the diffeomorphism constraint is solved, then the resulting equations of motions may turn out not be manifestly ultralocal; this appears for example in the spherically symmetric reduction of the theory \cite{Husain:2015cwa}, where the coordinate gauge is such that  the shift vector $N^a$ is not zero. But  if one chooses coordinate fixing conditions such that  $N^a=0$, then the dynamics remains  manifestly ultralocal. Thus whether or not one has manifest ultralocality depends on coordinate gauge.

\section{Linearized theory}

In this section we analyze the linearized theory of perturbations about a solution of the Hamiltonian equations of motion. We perturb around the flat background solution $q_{ab}=e_{ab}$, $\tilde{\pi}^{ab}=0$ and $N^a=0$, and  write the perturbed fields as
\be
q_{ab} = e_{ab} + h_{ab}, \qquad \qquad \tilde{\pi}^{ab} = 0+ \t{p}^{ab}, \qquad \qquad N^a = 0+ n^a.
\ee
Substituting this into the diffeomorphism constraint and equations of motion gives to first order  
\be
\label{lin-diff}
\nabla_a \t{\pi}^{ab} &\approx& \p_a \t{p}^{ab}  =0,\\
\label{lin-eom}
\dot{h}_{ab} &=& 2   \left(\t{p}_{ab} - \t{p} e_{ab} \right) + 2\p_{(a}n_{b)}, \\
\dot{\t{p}}^{ab} &=& 0.
 \ee
We note from these equations that any solution $\t{p}^{ab}_s$ of the diffeomorphism constraint gives a static source for the metric perturbation, which then evolves linearly in dust time (up to the diffeomorphism term if $n^a\ne 0$).
 


Our goal is to solve the diffeomorphism constraint and study the linearized equations of motion for a set of  physical variables. It is easiest to work in $k$-space, as done for example in \cite{Langois:1993}, using the Fourier transformed fields
\be
\bar{h}_{ab}(t,k) &=& \frac{1}{2\pi} \int d^2x\  e^{-i k_c x^c} h_{ab}(t,x),  \\
\bar{{p}}^{ab}(t,k) &=& \frac{1}{2\pi} \int d^2x\  e^{-i k_c x^c} \t{p}^{ab}(t,x), \\
\bar{n}^a(t,k) &=& \frac{1}{2\pi} \int d^2x\  e^{-i k_c x^c} n^a(t,x).
\ee
The transform of the symplectic term in the canonical action is 
\be
\int d^2x\  \dot{h}_{ab}(t,x) \t{p}^{ab}(t,x) &=& \frac{1}{(2 \pi)^2} \int d^2x\ d^2k \ d^2\bar{k}\  e^{i x^c (k_c+\bar{k}_c)}\dot{\bar{h}}_{ab}(t,k)\  \bar{p}^{ab}(t,\bar{k}) , \nonumber \\
&=& \int d^2k\  d^2\bar{k}\  \delta^2(k + \bar{k})\  \dot{\bar{h}}_{ab}(t,k) \bar{p}^{ab}(t,\bar{k}), \nonumber \\
&=& \int d^2k\  \dot{\bar{h}}_{ab}(t,k) \bar{p}^{ab}(t,-k).
\ee

The linearized  equations of motion in $k$-space become  
\be
\dot{\bar{h}}_{ab} &=& 2 \left( \bar{p}_{ab} - \bar{p} \delta_{ab} \right) + 2i k_{(a} \bar{n}_{b)},  \label{lin-k}\\
\dot{\bar{p}}^{ab} &=& 0.
\ee
The perturbations are still subject to the $k$-space diffeomorphism constraint,
\be
\bar{\mathcal{C}}^G (n) \equiv \bar{n}_a k_b \bar{p}^{ab} =0,  \label{kdiffeo}
\ee  
which contributes  the linear term in $\bar{n}^a$ to the $\dot{\bar{h}}_{ab}$ equation.
 
To fix the gauge and solve the diffeormorphism constraint, it is convenient to expand the  symmetric $k$-space tensors $(\bar{h}_{ab},  \bar{p}^{ab})$ in a suitable matrix basis  $A^I$, $I=1,2,3$, as
\be
\bar{h}_{ab}  = h_I  A^I_{ab}, \qquad  \qquad \bar{p}^{ab} = p^I A_I^{ab}.
\ee
 A convenient choice of basis is
\be
(A^1)_{ab}:= \frac{1}{\sqrt{2}} \delta_{ab}, \qquad (A^2)_{ab}:= \frac{\sqrt{2}}{|k|^2} k_a k_b - \frac{1}{\sqrt{2}} \delta_{ab}, \qquad
(A^3)_{ab}:= \frac{1}{\sqrt{2}|k|^2} \left( \epsilon_{ac}k_b k^c + \epsilon_{bc}k_a k^c \right) ,
\label{basis}
\ee
where $\epsilon_{ab}$ is the Levi-Civita symbol. These are defined in analogy with a similar basis used in 3+1 gravity: the first two correspond to scalar degrees of freedom, and the third  to a `vector' degree of freedom. This basis  is orthogonal and normalized with  the inner product
\be
(A^I, A^J) := (A^I)_{ab} (A^J)_{cd} \delta^{ac} \delta^{bd} = \delta^{IJ}.
\ee
Therefore the symplectic term becomes
\be
\dot{\bar{h}}_{ab} \bar{p}^{ab} = h_I A^I_{ab} p^J A_J^{ab} = h_I p^I,
\ee
so the expansion coefficients $h_I, p^I$ are canonically conjugate.  The basis also satisfies
\be
k^a A^1_{ab} = k^a A^2_{ab} = \frac{1}{\sqrt{2}} k_b, \qquad k^a A^3_{ab} = \frac{1}{\sqrt{2}} \epsilon_{bc} k^c , \qquad \delta^{ab} A^2_{ab} = \delta^{ab} A^3_{ab} = 0. \label{basiscond}
\ee
Thus $I=2,3$ are the traceless modes, and the linear combination $A^2-A^1$ is  transverse.  There is no transverse and traceless mode in this model since there are no metric perturbations satisfying both of these conditions:  the only solution to the two equations
\be
\delta^{ab} h_I A^I_{ab} = 0=k^a h_I A^I_{ab}
\ee
is $h_I = 0\ \forall \  I$.

We now write the linearized $k$-space equations in the basis (\ref{basis}). The diffeomorphism constraint takes a useful form: decomposing the first order shift vector in components parallel and perpendicular to $k^a$,
\be
\bar{n}^a = n_{\parallel} \frac{k^a}{|k|} + n_{\perp} \epsilon^{ab} \frac{k_b}{|k|},
\ee
this constraint (\ref{kdiffeo}) may be written as
\be
\bar{\mathcal{C}}^G = \bar{\mathcal{C}}^G_\parallel + \bar{\mathcal{C}}^G_\perp = n_{\parallel} \frac{|k|}{\sqrt{2}} \left( {p}_1 + {p}_2 \right) + n_{\perp} \frac{|k|}{\sqrt{2}} \ p_3.
\ee
The  gives  two separate constraints on the three momenta:
\be
p_1 + p_2  \approx 0, \ \ \ \  p_3 \approx 0. \label{2diffeos}
\ee
The equations of motion are
 \be
\dot{h}_1 &=& -2 {p}_1 + \sqrt{2} |k| \ n_\parallel \\
\dot{h}_2 &=& 2 {p}_2 + \sqrt{2} |k| \ n_\parallel, \\
\dot{h}_3 &=& 2p_3 +  \sqrt{2} |k| \ n_\perp, \\
\dot{{p}}_I &=& 0 \ \forall\ I,
\ee
where we have redefined $n_\parallel$ and  $n_\perp$ to absorb the factor of $i$ in (\ref{lin-k}).
This  shows  the advantage of the chosen  basis -- the equations of motion are decoupled, and the diffeomorphism constraint breaks neatly into components parallel and perpendicular to $k^a$.

\subsection{Gauge fixing and physical degrees of freedom}

We now obtain the fully reduced theory of physical degrees of freedom by  imposing two gauge conditions and solving the two diffeomorphism constraints (\ref{2diffeos}).  This will give a Hamiltonian theory of one pair of phase space  variables. There is more than one way to do this, but we will focus on the  choices that are natural in our chosen decomposition.

Given the form of the diffeomorphism constraint, it is natural to set the gauge $h_3=0$. This is second class with the constraint  $p_3\approx 0$. Requiring that the gauge condition be preserved in time means $\dot{h}_3 = 0$, which gives the condition  $n_\perp = 0$ on the linear shift.  These steps constitute a partial gauge fixing of the theory leaving the pair $(h_1,p_1)$ and $(h_2,p_2)$, and the remaining diffeomorphism constraint 
\be 
p_1+p_2 \approx 0. \label{reduceddiffeo}
\ee

The next step is to fix the remaining gauge freedom. There are multiple  ways of doing this, and in the following we present a few examples.

\subsubsection{Traceless gauge}
Let us recall from (\ref{basiscond}) that only the traceless mode remains if we set the gauge  $h_1 = 0$. This is second class with the reduced diffeomorphism constraint (\ref{reduceddiffeo}), which is solved by setting $p_1 = - p_2$. Dynamical preservation of this gauge  requires  $\dot{h}_1 = 0$, which gives
\be
n_\parallel = -\frac{\sqrt{2}}{|k|}\  p_2.
\ee
Thus the  fully gauge fixed theory has only the traceless mode $(h_2, p_2)$, satisfying the  equations of motion
\be
\dot{h}_2 = \dot{p}_2 = 0.
\ee
Notice the linear-order shift vector $n_\parallel$ works to cancel out the right hand side of the equation of motion for $h_2$.

Since the remaining degrees of freedom are static, the general solution is given by arbitrary functions of $k$:
\be
h_2 = \alpha(k), \qquad  p_2 = \beta(k) .
\ee

\subsubsection{Transverse gauge}
To leave the transverse mode as the remaining phase space pair, we use the gauge condition $h_1 + h_2 = 0$. This is second class with $\mathcal{C}^G_\parallel \approx 0$, which is again solved by setting  $p_1 = - p_2$. Keeping this fixed dynamically $\dot{h}_1 + \dot{h}_2= 0$ implies
\be
n_\parallel = \frac{\sqrt{2}}{|k|} p_1.
\ee
We can write the fully reduced theory using the  variables $(h_1, p_1)$. This gives  the same equations of motion as found in the traceless gauge:
\be
\dot{h}_1 = \dot{p}_1 = 0,
\ee
due to a cancellation coming from the solution to $n_\parallel$.

The solution to the equations of motion is again given in terms of arbitrary functions that are constant in time:
 \be
h_1 = \alpha(k), \qquad p_1 = \beta(k).
\ee

\subsubsection{Gauges with time dependence}
The two gauges used to fix diffeomorphism invariance discussed above give rise to the simplest of evolution equations. It is interesting to look at other  gauge choices, most of which  give non-trivial dynamical equations.  Let us consider a more general gauge  
\be
h_1 = f(t,k, h_2, p_1,p_2), \label{gengauge}
\ee
where $f$ is an arbitrary function. This is second class with  (\ref{reduceddiffeo}), which gives $p_1=-p_2$.  Dynamical preservation of this gauge now requires $\dot{h}_1 - \dot{f} = 0$. Using the equations of motion, this fixes the lapse to be 
\be
n_\parallel = - \frac{\sqrt{2} p_2}{k}  + \frac{1}{\sqrt{2}k}\  \frac{\p f}{\p t}\left(   1- \frac{\p f}{\p h_2}    \right)^{-1}.   
\ee
The resulting equations of motion for the physical  degrees of freedom  $(h_2,p_2)$ are  
\be
\dot{h}_2&=&   \frac{\p f}{\p t}\left(   1- \frac{\p f}{\p h_2}    \right)^{-1},     \\
\dot{p}_2&=& 0.  
\ee
The special case  $f= \mu t p_2$, $\mu=$constant,  reduces the first equation to that of a free particle $\dot{h}_2 =  \mu p_2$.

That different gauges give rise to very different dynamics is of course expected for a generally covariant theory. The interesting  feature is that the choice of dust time, together with  either the transverse or traceless gauges for the diffeomorphism constraint leads to the simplest  solution.

\subsubsection{Spacetime fields}
The solutions presented above are given in terms of $(h_I, p_I)$, which are matrix expansion coefficients in $k$-space. The first order ADM variables are obtained from inverse Fourier transforms. For example, the metric perturbation in the traceless gauge is  
\be
h_{ab}(x) 
= -\frac{1}{\sqrt{2}\pi}\partial_a \partial_b \int d^2k\  e^{ik_c x^c} \frac{\alpha(k)}{|k|^2} -
 \frac{e_{ab}}{2 \sqrt{2} \pi} \int d^2k\  e^{ik_c x^c} \alpha(k).
\ee
The other Fourier transforms take a similar form, and since each contains the arbitrary function $f$ for the general gauge,  one obtains a large  class of spacetimes for the linearized theory.

\section{Discussion}

We studied $3$-dimensional gravity coupled to dust in the dust time gauge. The structure of the canonical theory is similar to that in four dimensions, where the physical Hamiltonian is the same expression as the Hamiltonian constraint. We found that  the theory is ultralocal, a result peculiar to three spacetime dimensions  due to the evolution equation for $\pi^{ab}$.  For an understanding of the local metric degree of freedom in this time gauge, we analyzed the  linearized theory about flat spacetime in $k$-space. This gives a curious result for the transverse or traceless gauge: the linearized evolution equations  are trivial, which means that any initial perturbations do not evolve, but remain frozen.   

This result means that the fully gauge fixed linearized theory in dust time gauge, in the chosen $k$-space basis, is such that that the physical Hamiltonian of the physical modes is exactly zero.  The simplicity of this solution is due to the choice of time gauge, the matrix basis, and the transverse or traceless coordinate gauge. The situation is analogous to the classical mechanics problem of finding the time dependent canonical transformation that maps a non-trivial hamiltonian to the trivial one.  Had we used another time gauge at the outset, the physical Hamiltonian and linearized equations of motion would have been very different, and non-trivial, because the solution of the Hamiltonian constraint would not be as simple as the one provided by the dust time gauge. We see this even within the dust time gauge, if the remaining gauge choices are made in a more general way, as in (\ref{gengauge}). It remains to explore the full non-linear ultralocal equations 
(\ref{3d1}) and  (\ref{3d2}).

Beyond 3d, little is known about the canonical quantization of a matter--gravity system in 3+1 dimensions. It is natural to apply the procedure used here to the linearized theory for 3+1 gravity coupled to dust, a work presently in progress. It is of interest to see whether the simplifications from the dust time gauge and choice of matrix basis in $k$-space produce  new insights in this difficult problem. Another direction for further work is the inclusion of  other matter fields in addition to dust; the physical Hamiltonian for such cases is a simple sum of the gravity Hamiltonian ${\cal H}_G$ and  the standard hamiltonian for the matter field \cite{Husain:2011tk}.

\begin{acknowledgments}

This work was supported by NSERC of Canada, and an AARMS Postdoctoral Fellowship to J.Z.

\end{acknowledgments}

\bibliography{3dDust}

\end{document}